# Impact load mitigation in sandwich beams using local resonators

B. Sharma and C.T. Sun[1]

School of Aeronautics and Astronautics, Purdue University, Armstrong Hall,

West Lafayette, Indiana 47907 USA

**Abstract**

Dynamic response of sandwich beams with resonators embedded in the cores subjected to impact loads is studied. Using finite element models the effectiveness of various local resonator frequencies under a given impact load is compared to the behavior of an equivalent mass beam. It is shown that addition of appropriately chosen local resonators into the sandwich beam is an effective method of improving its flexural bending behavior under impact loads. The effect of a given local resonance frequency under different impact load durations is also studied. It is demonstrated that the choice of appropriate local resonance frequency depends on the impact duration. Further, by performing transverse impact experiments, the finite element models are verified and the advantage of using internal resonators under impact loading conditions is demonstrated.

---

[1] Corresponding author. Tel.: +1 765 494 5130; fax: +1 765 494 0307.

  *E-mail address*: sun@purdue.edu (C.T. Sun).

**Introduction**

Sandwich structures made by bonding high stiffness facesheets to a low density core are finding increasing use due to their highly attractive strength to weight ratio. However, their major drawback is their poor performance under impact loads and dynamic conditions [1], [2], [3], [4]. Thus, to further encourage adoption of sandwich structures as a viable design option it is important to understand and improve the behavior of such structures under various dynamic conditions.

The behavior of sandwich structures under impact loading has been investigated by a number of researchers and excellent reviews can be found in [2] and [5]. The majority of the research focus has been on characterizing damage induced due to impact loading, though some researchers have also focused on energy absorption characteristics of sandwich structures. In a weight-for-weight comparison the energy absorption capabilities of sandwich panels have been found to be superior to those of metals [6], [7]. It has also been shown that the energy absorption capacity of sandwich panels increases with increasing transverse stiffness and increasing impact velocity [8], [9]. Using through-thickness split Hopkinson pressure bar tests, Mahfuz et al. [10] showed that the core material absorbs most of the impact energy. To improve the energy absorption capacity of sandwich structures, Kenny and Torre [11] proposed a corrugated sandwich panel and showed that such a design offers a superior energy absorption capacity compared with traditional sandwich structures.

Recently, Chen and Sun [12] demonstrated that addition of resonators to sandwich cores is an effective method of improving the wave attenuation behavior of sandwich beams. The local resonance behavior of the inserted resonators was shown to induce a wave attenuation bandgap which allows for effective attenuation of harmonic flexural waves. In this study, the effectiveness of such a sandwich design under impact loads is considered. Impact loads are typically short duration loads and are thus broad spectrum in nature, as opposed to single frequency harmonic loads. The purpose of this study is to study the effectiveness of local resonators in attenuating such broad spectrum loads. Finite element models are used to evaluate the response of sandwich beams with and without resonators. The effectiveness of different resonators under a prescribed impact load is studied. The performance of a chosen internal resonator under different impact loads is also considered. Finally, transverse impact experiments are performed to verify the finite element models and to experimentally demonstrate the advantage of using local resonators.

**Sandwich Beam Construction**

A typical sandwich beam with internal resonators is shown in Figure 1. Typically, the bending stiffness of the face sheets greatly exceeds the stiffness of the core, while the core shear stiffness is significantly greater than the shear stiffness offered by the thin face sheets. Thus, it can be safely assumed that the bending behavior of the sandwich beam is entirely controlled by the facesheets, while the shear behavior is controlled by the foam core. The resonator is a spring-mass element contained in a cylindrical plastic casing which are embedded in the sandwich core.

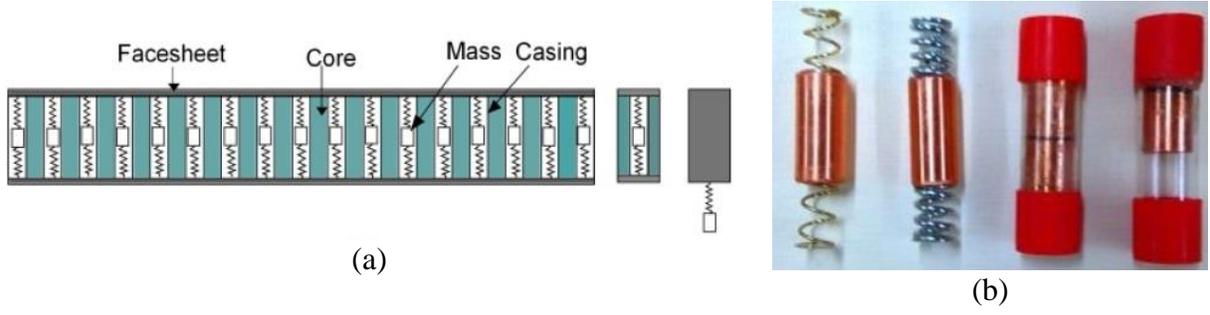

**Figure 1.** (a) Left to right: Sandwich beam with internal resonators, a unit cell and an equivalent Timoshenko representation, (b) Left to right: 100 Hz resonator, 300 Hz resonator, encased resonator, equivalent mass element.

Such a sandwich beam can be modeled as a solid Timoshenko beam with the resonators attached to it [12] as shown in Figure 1(a). The equations of motion of this beam are:

$$EI\varphi'' - GA(v' + \varphi) - \rho I \ddot{\varphi} = 0 \tag{1}$$

$$GA(v'' + \varphi') - \rho A \ddot{v} + \frac{2k}{a}(v_1 - v) = 0 \tag{2}$$

$$\frac{2k}{a}(v_1 - v) + \frac{m}{a}\ddot{v_1} = 0 \tag{3}$$

where $v$ is vertical displacement of the beam, $v_1$ is the vertical displacement of the resonator mass, $\varphi$ is rotation of the cross-section, $EI$ is bending stiffness contributed by the facesheets, $GA$ is transverse shear rigidity contributed by the core, and $\rho A$ and $\rho I$ are transverse and rotary inertias, respectively, k is the resonator spring stiffness, m is the resonator mass, and a is the unit cell length. Assuming harmonic wave propagation, the dispersion equation for such a beam is found to be of the form:

$$A_1\omega^6 + A_2\omega^4 + A_3\omega^2 + A_4 = 0 \tag{4}$$

where

$$A_1 = -\rho I \rho A \left(\frac{m}{a}\right) \tag{5}$$

$$A_2 = \left\{(\rho IGA + EI\rho A)\left(\frac{m}{a}\right)q^2 + GA\rho A\left(\frac{m}{a}\right) + \rho I\rho A\left(\frac{2k}{a}\right) + \left(\frac{2km}{a}\right)\rho I\right\} \quad (6)$$

$$A_3 = -\left\{EIGA\left(\frac{m}{a}\right)q^4 + \left\{(\rho IGA + EI\rho A\left(\frac{2k}{a}\right) + \left(\frac{2km}{a^2}\right)EI\right)\right\}q^2 + GA\rho A\left(\frac{2k}{a}\right) + \left(\frac{2km}{a^2}\right)GA\right\} \quad (7)$$

$$A_4 = EIGA\left(\frac{2k}{a}\right)q^4 \quad (8)$$

The wave propagation behavior of the beam can be analyzed by using Equations 4-8 to obtain the frequency, $\omega$ as a function of the wave number, q.

**Numerical Analysis**

*Dispersion curves and bandgap*

Dispersion curves for a sandwich beam with resonators embedded in the sandwich core are obtained using Equation 4. The sandwich beam properties used to obtain the dispersion behavior are listed in Table 1. The resonator mass is assumed as 23 g while the spring constant is assumed as 326881 N/m, giving a local resonance frequency of 600 Hz.

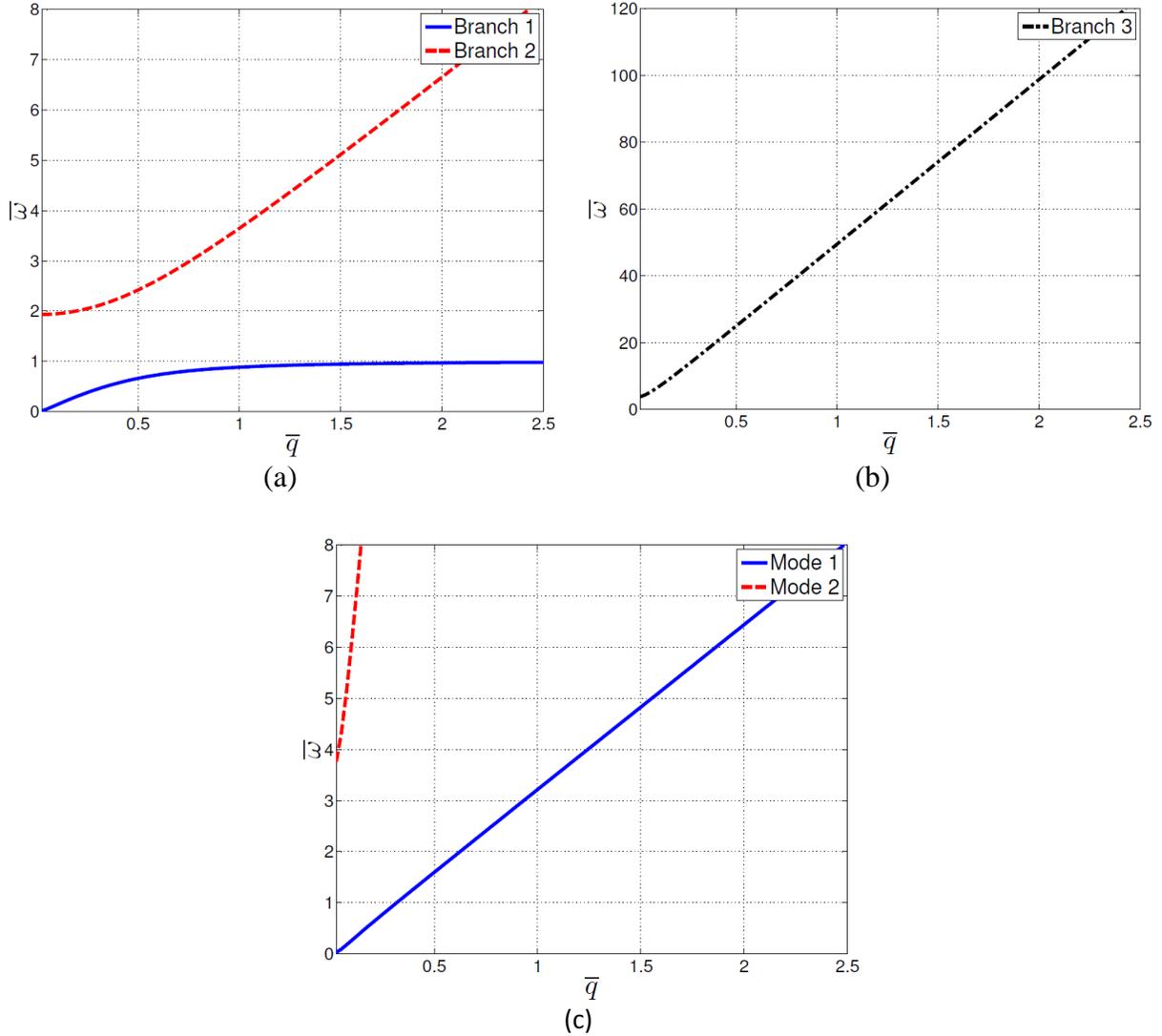

**Figure 2.** (a - b) Dispersion curves for sandwich beam with resonators, (c) Dispersion curves for sandwich beam without resonators.

Figures 2 (a-b) show the dispersion curves for a sandwich beam with resonators tuned at 600 Hz. Note that the solution provides three branches since the addition of the resonators provides an additional degree of freedom. For the purpose of comparison, the dispersion curves for the same sandwich beam without resonators are also shown in Figure 2(c). The normalized frequency is defined as $\bar{\omega} = \omega/\omega_o$, where $\omega_o = \sqrt{k/m}$ is the local resonance frequency, and the normalized wave number is defined as $\bar{q} = qa$, where a is the unit cell length. It can be seen that the addition of resonators causes a bandgap to appear around the local resonance frequency, which is absent in the beam without the resonators. For the parameters chosen, the bandgap is found to exist from $\bar{\omega}$ = 1 to 1.9. The effect of various parameters on the bandgap width is studied in detail in [12].

Finite element simulations are performed to verify the existence of this wave attenuation bandgap. Sandwich beams with resonators uniformly distributed along the beam length are modeled with a unit periodic displacement applied at the center of the beam. To capture the propagating wave effects and avoid reflections, a 100 m beam is modeled with symmetry conditions applied at the left end and the right end kept free, thus effectively simulating a 200 m long centrally loaded free-free beam. To understand the behavior of the beam at various frequencies, input displacements are applied at a frequency below the bandgap starting frequency (400 Hz), two frequencies lying inside the bandgap (650 Hz and 800 Hz), and a frequency lying outside the bandgap frequency (1400 Hz).

The output displacements are obtained 1 m away from input. As mentioned earlier, the bandgap exists from 600 Hz till 1140 Hz. For 400 Hz and 1400 Hz, frequencies lying outside the bandgap, magnitude of the output displacements are equal to the input displacements and no wave attenuation is observed. However, for frequencies lying inside the bandgap, 650 Hz and 800 Hz, significant attenuation is obtained as predicted by the dispersion curves shown in Figure 3. Thus, addition of resonators causes a wave attenuation bandgap to occur. The rest of this paper deals with the use of such a wave attenuation bandgap to improve performance of sandwich beams under impact loads.

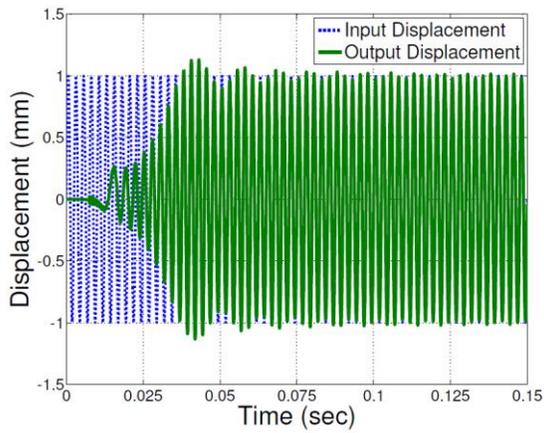
(a)

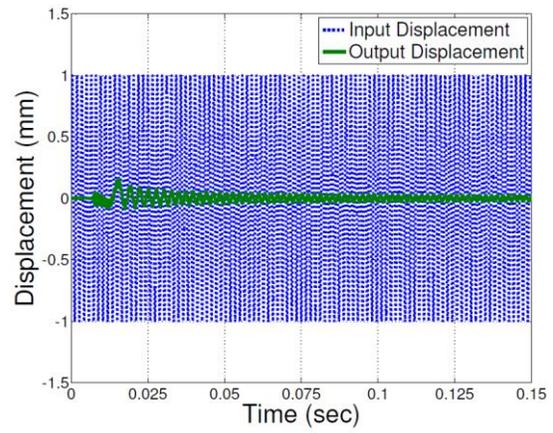
(b)

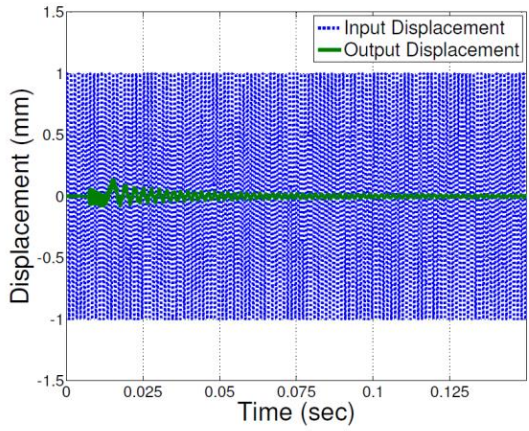
(c)

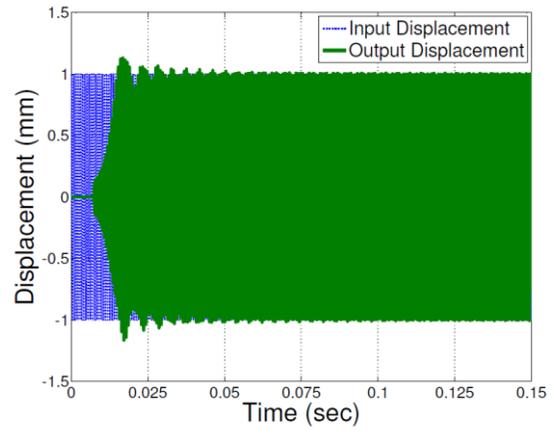
(d)

**Figure 3.** Input and output displacements for sandwich beam with resonators tuned at 600 Hz and excited at: (a) 400 Hz, (b) 650 Hz, (c) 800 Hz, and (d) 1400 Hz.

*Simulation of impact response*

Finite element simulations are performed to investigate the effect of internal resonators in sandwich beams subjected to impact loads. Sandwich beams with resonators uniformly distributed along the beam length are modeled with the impact load applied at the center of the beam. To capture the propagating wave effects and avoid reflections, a 100 m beam is modeled with symmetry conditions applied at the left end and the right end kept free, thus effectively simulating a 200 m long centrally loaded free-free beam as shown in Figure 4.

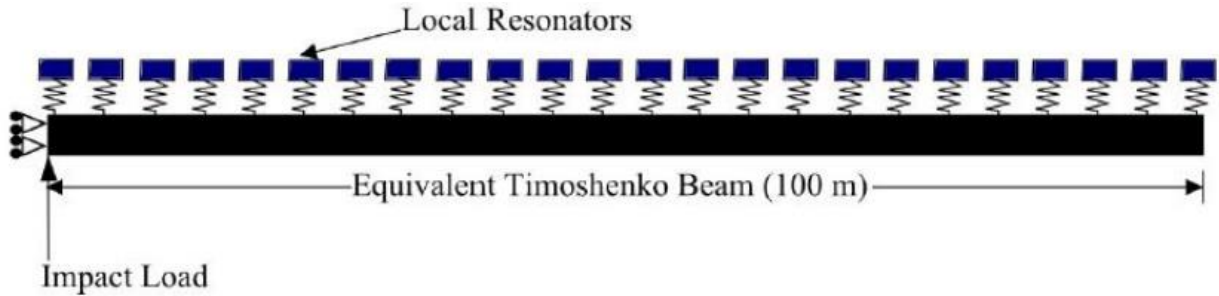

**Figure 4.** Finite element model schematic. Note that point masses are used as the resonator mass.

The Timoshenko beam theory is used to model the effective sandwich beam [14]. The material constants and the sandwich section dimensions are given in Table 1, where $E_f$ is the facesheet elastic modulus, $h_f$ and $h_c$ are the facesheet and core thicknesses respectively, b is the sandwich width, a is the distance between individual resonators, G is the core shear modulus, and $\rho_f$ and $\rho_c$ are the facesheet and core densities, respectively. Planar Timoshenko beam elements with linear interpolation (B21) provided by commercial code Abaqus are used to model the sandwich beam. The effective sandwich beam section properties are calculated as described in [14]. Each element is assigned a general beam section with bending rigidity EI, shear rigidity GA, rotary inertia $\rho I$, and mass per unit length $\rho A$, as given in Table 2. A unit cell length of 25 mm is chosen and five elements are used to model each unit cell. The resonators are discretely attached to the beam element and the resonator masses are modeled as equivalent point masses. To avoid round-off errors due to large number of increments, linear explicit analysis with double precision is carried out using Abaqus/Explicit. It should be noted that no damping is assumed in any of the numerical simulations.

**Table 1.** Material constants and dimensions of sandwich beam components

| $E_f$ (GPa) | $h_f$ (m) | $h_c$ (m) | b(m) | a(m) | G(Mpa) | $\rho_f$(Kgm$^{-3}$) | $\rho_c$(Kgm$^{-3}$) |
|---|---|---|---|---|---|---|---|
| 68.9 | 1.588e-3 | 0.0508 | 0.0254 | 0.025 | 20.0 | 2700 | 64 |

**Table 2.** Sandwich beam effective properties

| EI (Pa m$^4$) | GA (Pa m$^2$) | $\rho A$(kg/m) | $\rho I$(kg m) |
|---|---|---|---|
| 3750.1 | 25400 | 0.2957 | 1.6444e-4 |

The effectiveness of resonators in attenuating the impact load is assessed by comparing it with a sandwich beam of equivalent mass. Such a beam can be modeled in two different ways. The first method is to add the mass of the resonators to the mass per unit length assigned to the beam element, while the second method is to replace the resonators with equivalent point masses. For the purpose of this analysis, the latter method is chosen due to the added advantage of being able to replicate the periodicity of the local resonators. The unit cell length, beam material properties, mesh size, and the boundary conditions are kept the same.

The effect of the local resonance frequency of the resonators on a given impact load is studied by subjecting the beam to an impact load and varying the local resonance frequency. The impact is modeled as a smooth triangular pulse of duration 1 ms. Local resonators are tuned to a specific resonance frequency by varying the spring stiffness while keeping the mass constant. The mass is kept constant at 23 g for all the cases analyzed in this study. The resonator frequencies analyzed for this case are 600 Hz, 1000 Hz and 2000 Hz. The associated spring constants are given in Table 3.

**Table 3.** Resonator spring constants

| Resonator Frequency (Hz) | 100 | 600 | 1000 | 2000 |
|---|---|---|---|---|
| Spring Constant (N/m) | 8290.46 | 326881 | 908003 | 3632014.42 |

The effect of a fixed local resonance frequency on impact loads of different durations is also analyzed. The resonators are tuned to a resonance frequency of 600 Hz and three impact loads of duration 4 ms, 2 ms, and 1 ms are studied. The impact durations are chosen so that the maximum frequencies associated with the impacts are 1000 Hz, 2000 Hz and 4000 Hz, respectively. The impact loads and their frequency spectrums are shown in Figure 5(a) and 5(b), respectively.

*Results and Discussion*

In order to understand the effect of different local resonance frequencies on a given impact load, three different resonator frequencies, 600 Hz, 1000 Hz and 2000 Hz, are analyzed.

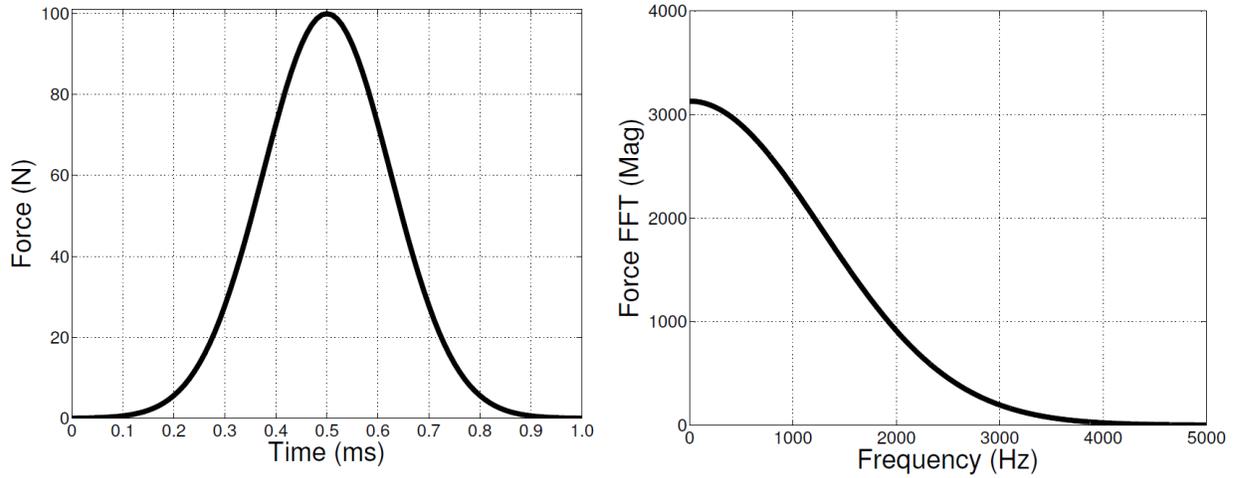

**Figure 5.** (a) Impact load history; (b) frequency spectrum of impact load.

Figure 6 shows the resultant bending strains in the beams as measured 2 m away from the impact location. It should be noted that the choice of beam length ensures that no reflections are present in the results shown. For comparison, the bending strain for an equivalent mass beam is also plotted.

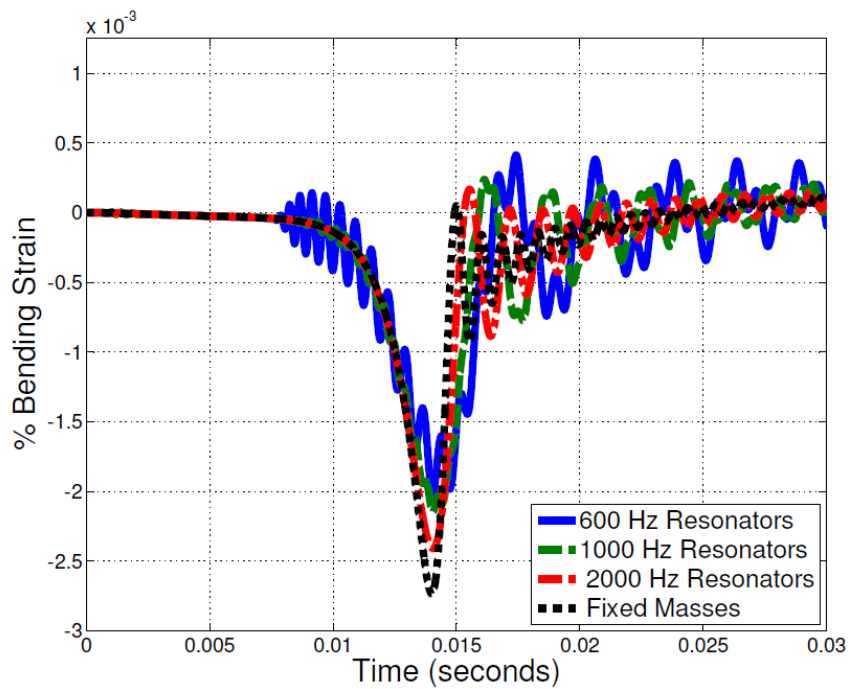

**Figure 6**. Bending strain comparison of beams with resonators tuned at different frequencies and beam with fixed mass.

The strain magnitudes in the beams with the resonators are lower as compared to the beam without resonators. This is due to the energy absorption associated with the motion of the resonator masses [13]. No such energy absorption mechanism is available for the beam without the resonators. Among the three resonators chosen, the 600 Hz resonators provide better attenuation than 1000 Hz and 2000 Hz. Since the system is dispersive and due to the difference in wave speeds of the systems, the percent reduction in amplitude is difficult to judge and only a qualitative conclusion can be accurately made. For the given impact load and as measured 2 m away from the impact location, the resonators tuned to 600 Hz offer the best performance by attenuating the wave by approximately 26%, followed by the 1000 Hz resonators at 21% and the 2000 Hz resonators at about 11%. This can be explained by looking at the frequency transform of the measured strains.

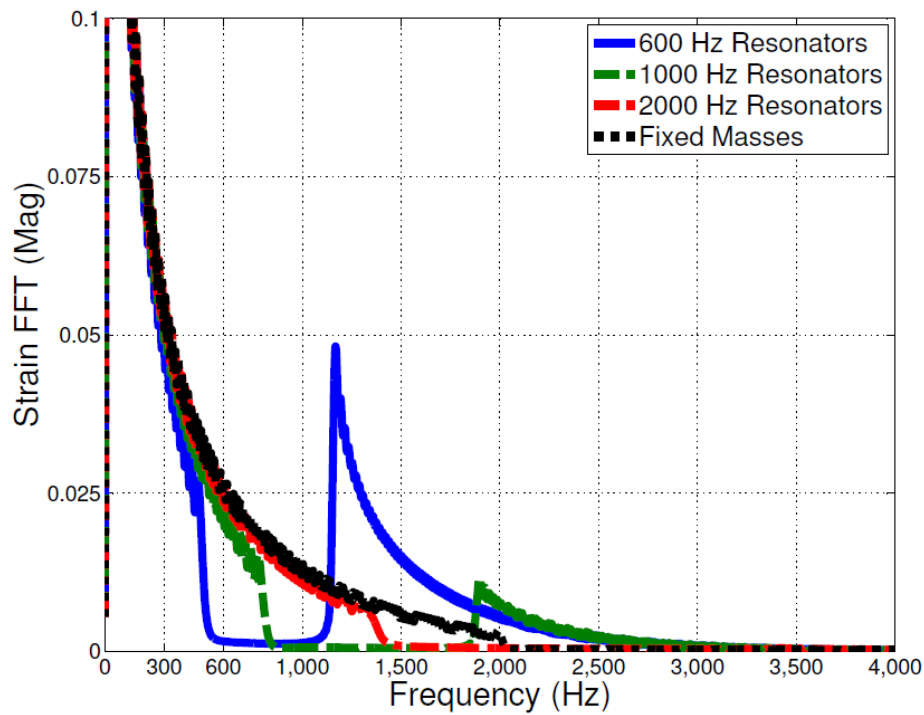

**Figure 7.** Frequency spectrum of bending strains as measured 2 m away from point of impact.

Figure 7 shows the frequency transform of the strains shown in Figure 6. For a typical impact load, as seen from the frequency spectrum of the input wave, the lower frequency content is considerably greater than the higher frequency content. The bandgaps for the individual resonators can be clearly seen in Figure 7. The 600 Hz resonators attenuate the waves between 535-1131 Hz; the 1000 Hz resonators create a bandgap between 852-1872 Hz; while the 2000 Hz resonators attenuate the waves above 1422 Hz. Thus, due to the nature of the impact load, though the higher frequency resonators offer larger bandgaps, the lower frequency resonator offers a much better performance by attenuating more of the lower frequency content. However, it should be noted that the bandgap width reduces considerably as the resonance frequency decreases [14], and thus simply tuning the resonators to a lower frequency does not give the optimum wave attenuation characteristics. This can be seen in Figure 8 where the strains for beams with 600 Hz and 100 Hz resonators are compared. The 600 Hz resonators clearly reduce the strain more effectively than the 100 Hz resonators. Thus, there exists a particular optimum local resonance frequency which would give the best performance, which may be found using an optimization routine. This optimization was not taken up in this study.

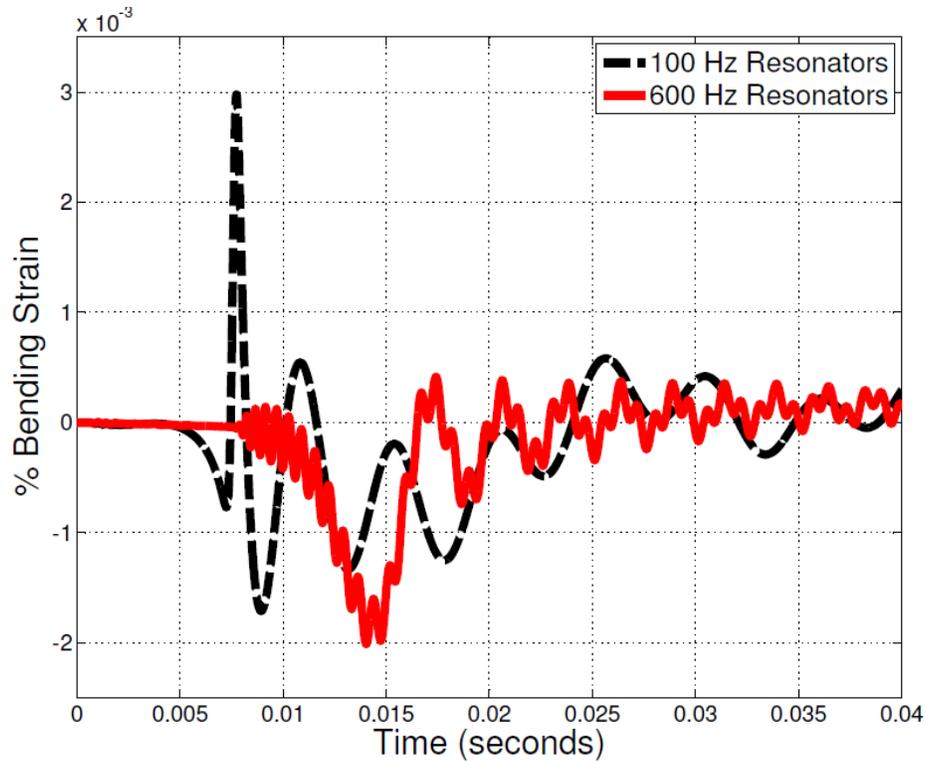

**Figure 8.** Bending strain comparison for beams with resonators tuned at 100 Hz and 600 Hz, respectively.

The effectiveness of a given resonator frequency under different impact loads is also studied. For the purpose of this study, the resonators are tuned at 600Hz and their effectiveness under three different impact loads is analyzed. The chosen impact loads and their frequency transforms are shown in Figure 9(a). It should be noted that the magnitude of the impact force is kept constant for all three cases, but due to the difference in their durations the energy content of the three pulses is different as can be seen from Figure 9(b).

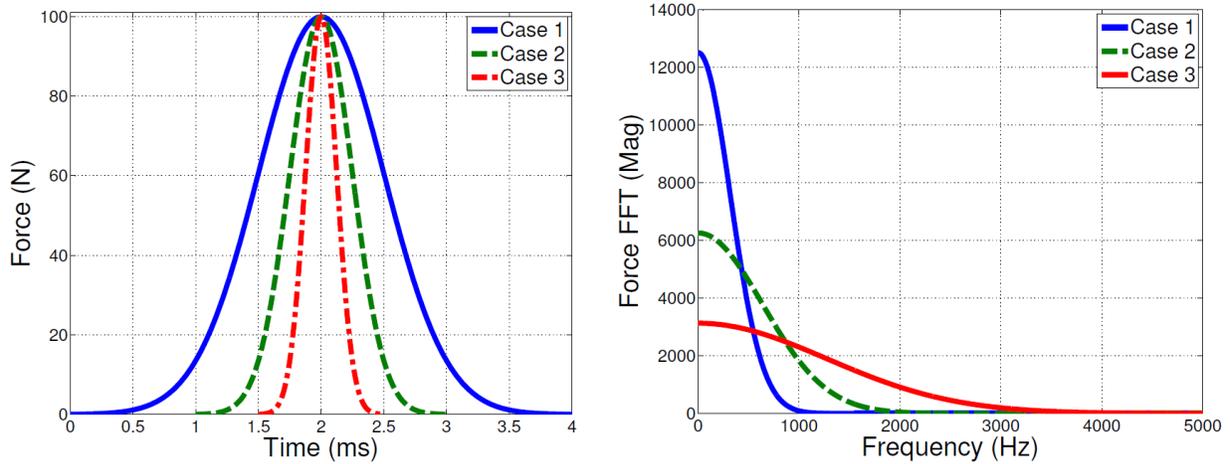

**Figure 9.** (a) Comparison of impact force histories; (b) comparison of frequency spectrum of the impact forces.

Figures 10, 11 and 12 show the strains measured for the sandwich beams subjected to impact loads of duration 4 ms, 2 ms, and 1 ms, respectively. It can be seen that the 600 Hz resonators are more effective in attenuating loads with higher frequency content. This can be explained by looking at the frequency spectrum of the input load. From the previous analysis, it is known that the wave attenuation bandgap created due to the 600 Hz resonators extends from 535 Hz to 1131 Hz. Thus, the chosen resonators attenuate a wider spectrum of frequencies for pulses of duration 1 ms and 2 ms as compared to a 4 ms pulse. Thus, while choosing a resonator frequency it is also important to consider the typical impact load durations expected to be encountered by the structure.

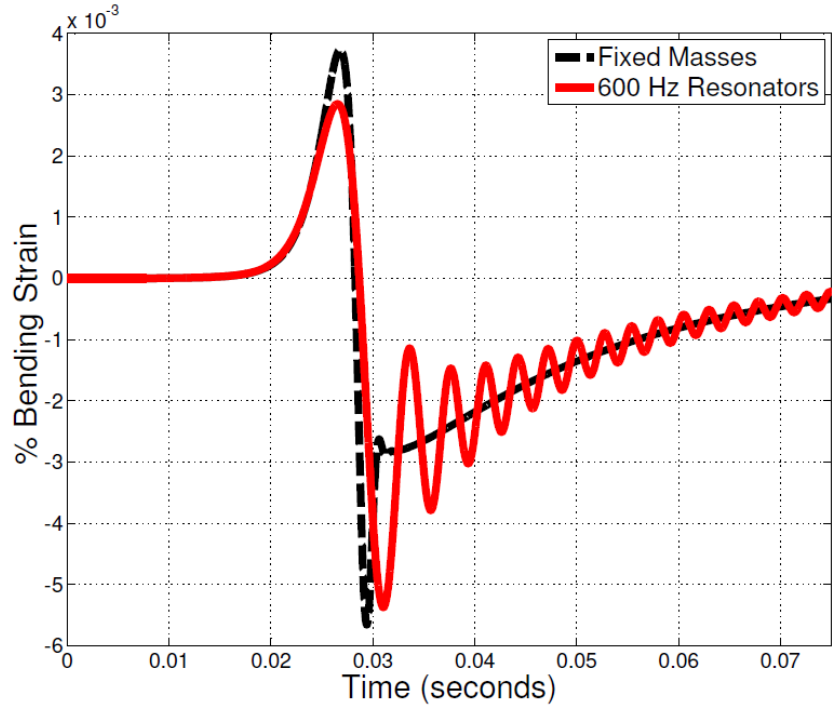

**Figure 10.** Comparison of bending strains for sandwich beams subjected to an impact load of duration 4 ms.

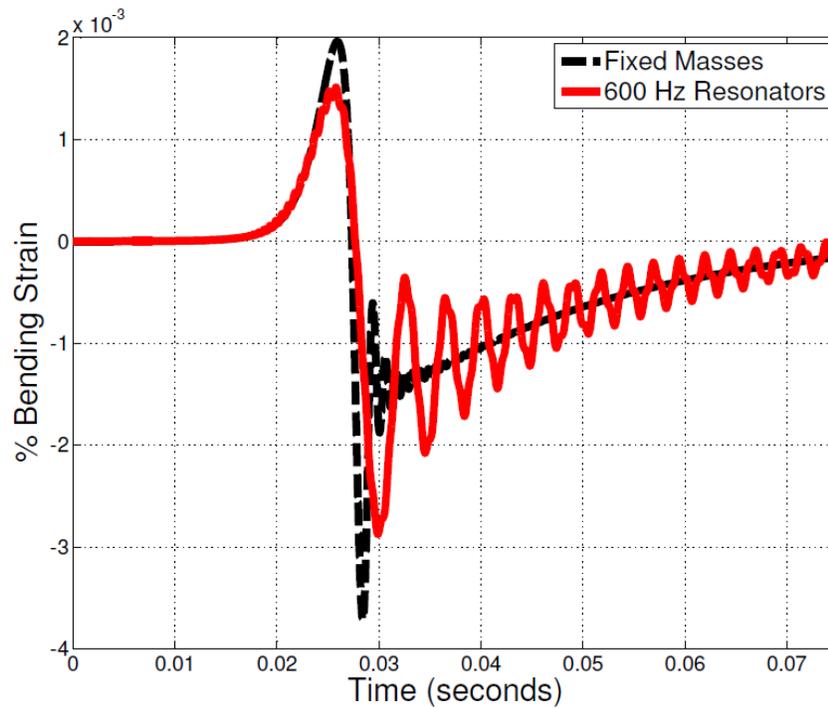

**Figure 11.** Comparison of bending strains for sandwich beams subjected to an impact load of duration 2 ms.

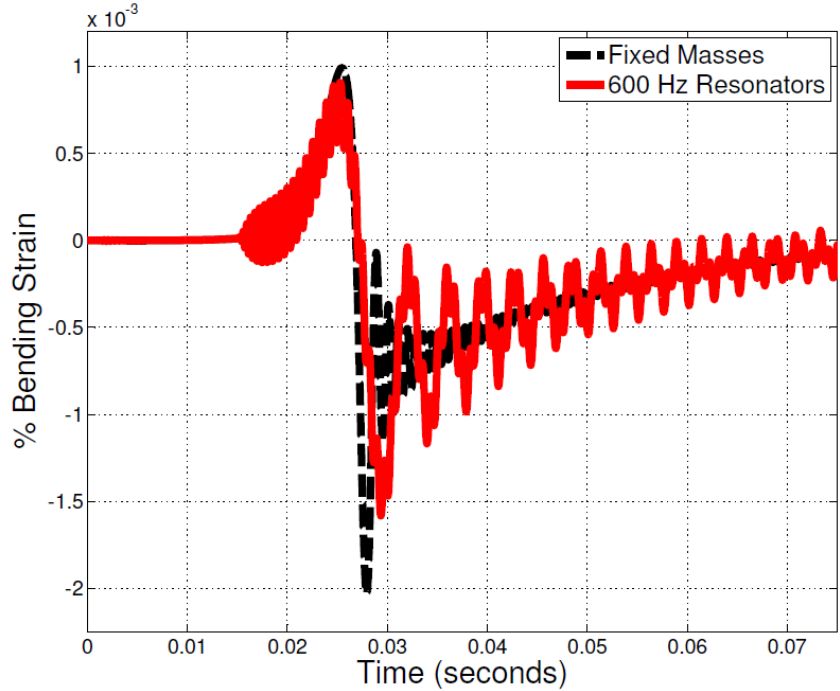

**Figure 22.** Comparison of bending strains for sandwich beams subjected to an impact load of duration 1 ms.

**Experiments**

*Experimental setup and method*

Impact experiments are performed to demonstrate the effectiveness of internal resonators for attenuating impact loads and to validate the finite element models used in the previous section. A 182.8 cm (72 in) long sandwich beam with rectangular cross-section of height 5.08 cm (2 in) and width 2.54 cm (1 in) is manufactured. The facesheets are made using Aluminum 6061 while the sandwich core consists of FR-7104 foam core obtained from General Plastics. Resonators were inserted into the core by drilling 71 holes through the thickness periodically. In order to allow their reuse, the resonators are inserted into plastic casings, which are in turn inserted into the drilled holes, after which the facesheets are bonded to the core using commercially available two part epoxy. The resonator mass consists of 2.54 cm (1 in) long copper rods, turned down using a lathe to a diameter of 1.06 cm (0.42 in), to obtain a final mass of 23 gm. Local resonance frequencies of 100 Hz and 300 Hz are obtained by bonding the mass to a spring of appropriate stiffness obtained from Century Spring Corp. In order to maximize the effectiveness of the resonators, they are inserted as a group, i.e., all the 100 Hz resonators are inserted into the left half of the beam, while the 300 Hz resonators are inserted into the right half of the beam. The effective shear rigidity of the sandwich beam with the embedded plastic casings is evaluated using a three point bending test [15]. The dimensions and the material properties are summarized in Tables 1 and 2.

To compare the results of the beam with resonators, two more beams are manufactured: one with the same total mass as the beam with resonators, and one without the added mass. For the beam with the same mass, the copper rods are bonded to the plastic caps of the resonator casings using epoxy and inserted periodically into the beam similar to the beam with resonators. For the beam without the added mass, empty resonator casings are inserted into beam core periodically. This is done to maintain the same effective shear rigidity for all three cases.

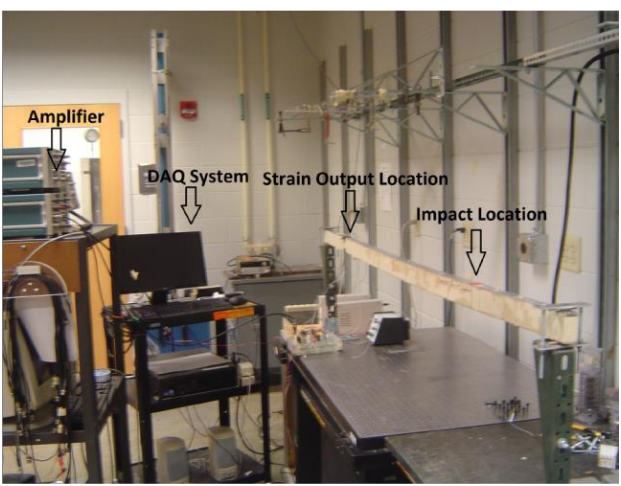

(a)

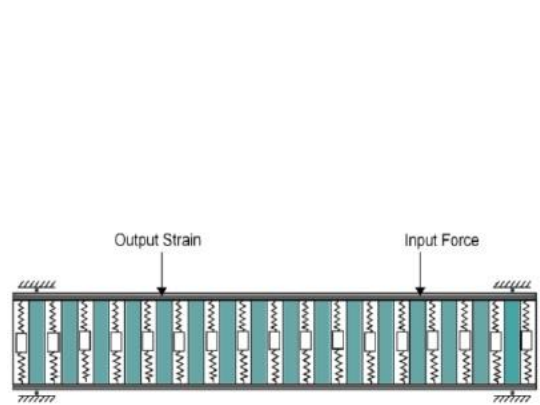

(b)

**Figure 13.** (a) Experimental setup; (b) schematic of force input and strain output locations.

Figure 13(a) shows the experimental set-up. The sandwich beam is simply supported at 2.5 cm (1 in) away from both ends. The input force is generated and measured by manually impacting the beam with a PCB 086C03 impact hammer. Figure 14 shows ten different impacts generated by this method. It can be seen that though the magnitude of the impact force varies in the impacts shown, the pulse shape and duration obtained for each impact are extremely close. The bending strains along the beam are measured using 1000 $\Omega$ strain gauges mounted in a balanced full bridge configuration to obtain the best strain resolution. In order to avoid a very short impact pulse, a 3.175 mm (0.125 in) thick rubber piece is stuck to the upper facesheet at the point of impact. All three beams are impacted 38 cm (15 in) away from the right-end while the strain signals are measured 1.0668 m (42 in) away from the impact point, as shown in Figure 13(b). The strain signals are amplified using an amplifier with a voltage gain of 500 and a low-pass filter tuned at 6400 Hz.

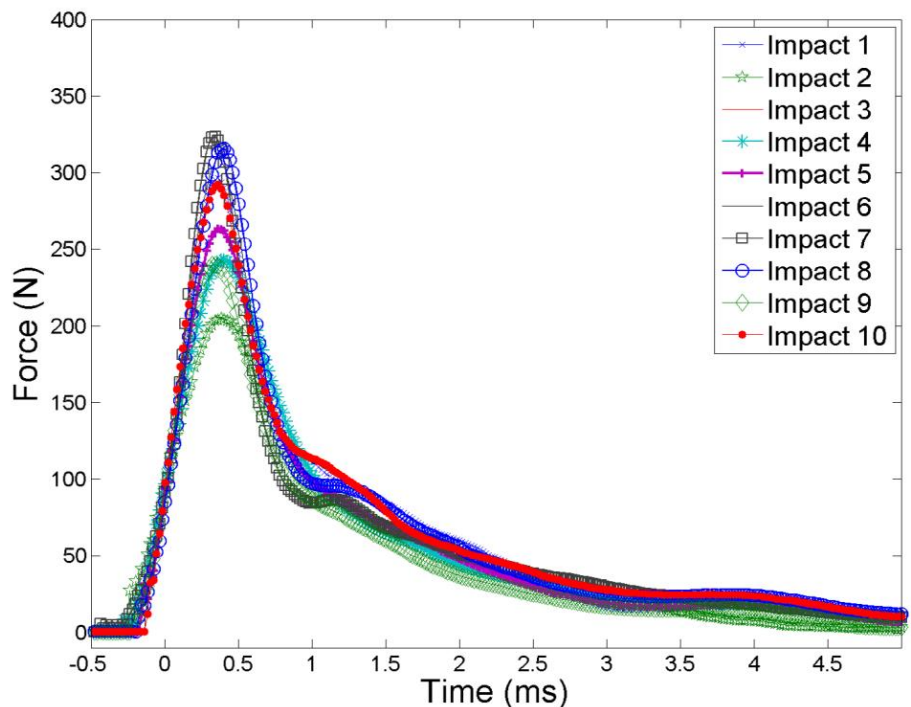

**Figure 34.** Comparison of different impact histories generated using impact hammer.

The experimental results are compared against results obtained from FE models. The sandwich beams used in the experiments are modeled as described in the previous section. The experimentally measured input force is used as the input for the FE models while the strains are measured at the same location as the experiments.

*Results and discussion*

Figure 15(a) shows the input force generated by the impact hammer on the three beams. A force of magnitude 340 N is generated for all three cases. However, different impact durations are obtained for the beam with resonators and the other two beams. The impact for the beam with resonators is approximately 4.2 ms long while for the other two cases a shorter pulse of 3.5 ms duration is generated. A comparison of their frequency spectrum is shown in Figure 15(b). It is seen that the magnitude of the frequency spectrum for the force generated due to impact with the beam with resonators is a bit higher for the lower frequencies (< 200 Hz) as compared to that for the other two beams. Thus, the input energy imparted to the beam with resonators is greater than that for the other two beams.

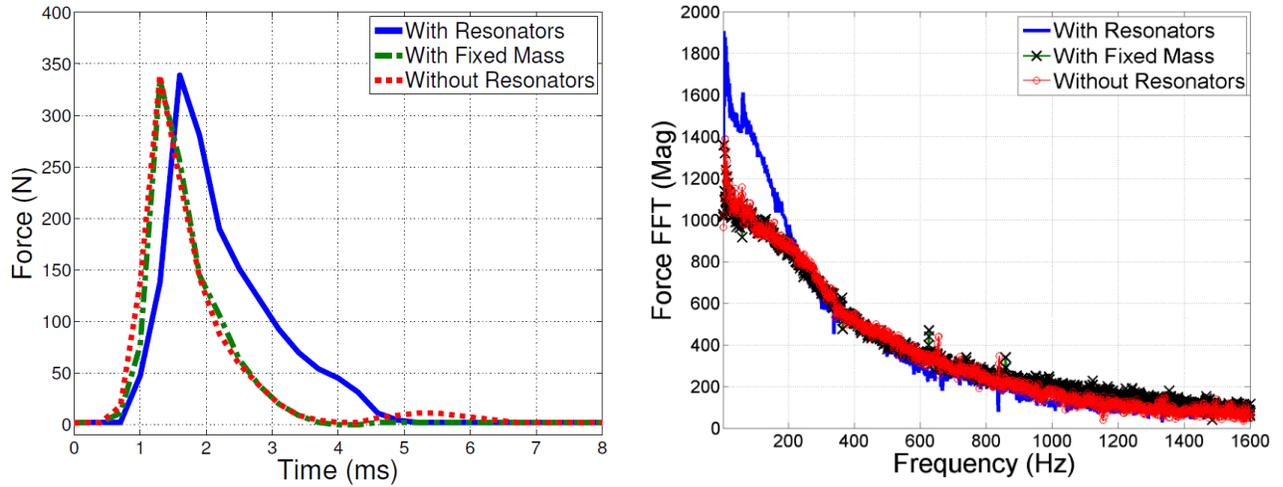

**Figure 15.** (a) Comparison of input forces generated on the three beams; (b) comparison of their input frequency spectrum.

The strains measured during the experiments are compared with the numerical results in Figure 16, 17 and 18. It should be noted that due to the length of the beam and the dispersive nature of the waves, there are reflections included in the measured results. For this discussion, we thus restrict ourselves to the initial part of the strain signal which has the least reflections included in it.

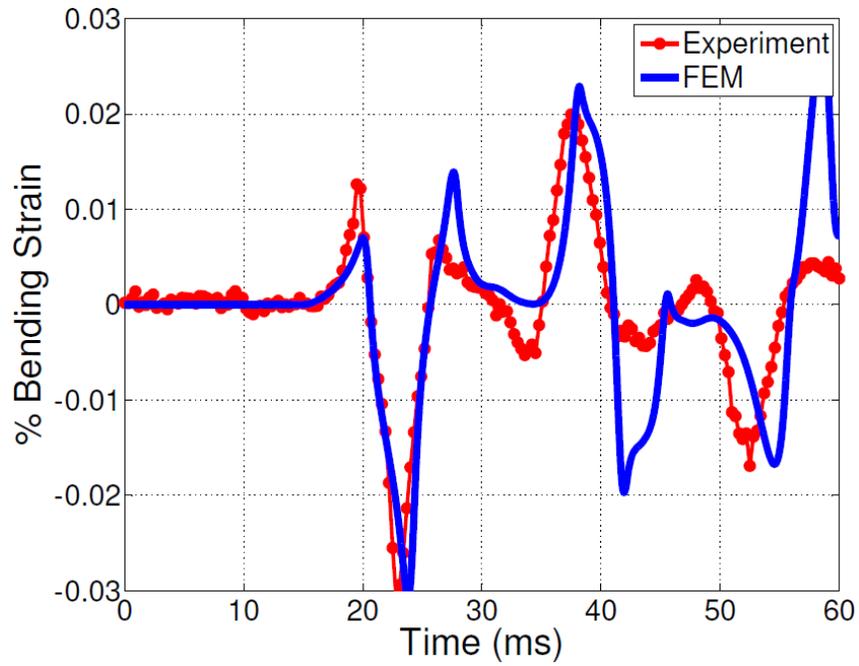

**Figure 16.** Comparison of experimental and numerical strains for beam without resonators.

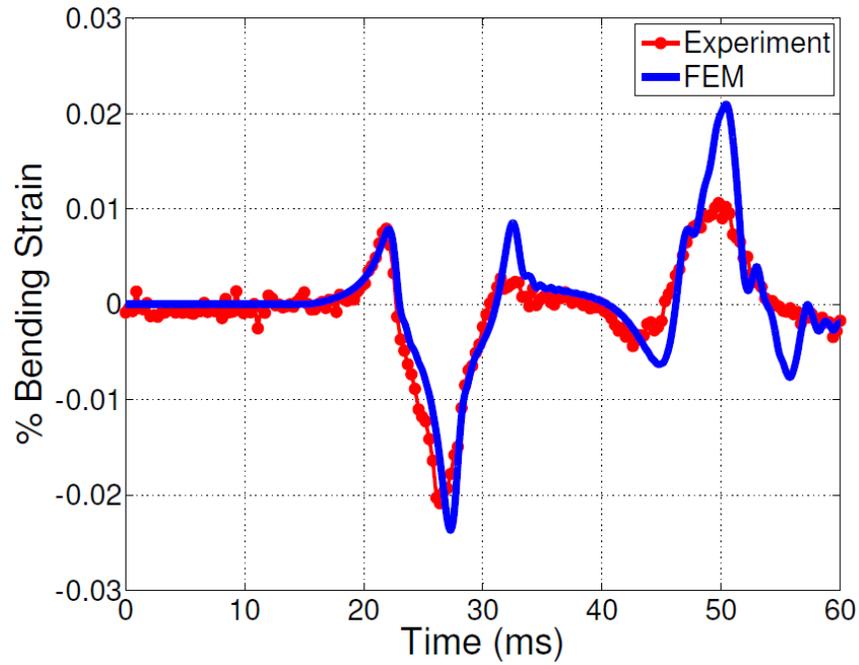

**Figure 17.** Comparison of experimental and numerical strains for beam with fixed mass.

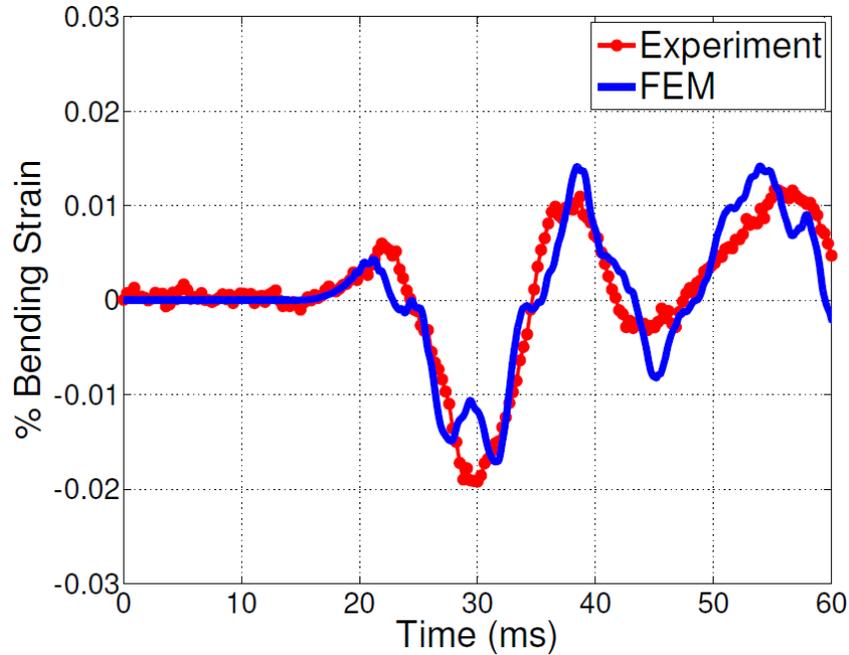

**Figure 18.** Comparison of experimental and numerical strains for beam with resonators.

A good match is observed between the experimental and numerical strains. However, some discrepancies are noticed between the two. An important difference between the numerical model and the experiments is that damping has been completely ignored in the numerical models. The effect of this assumption can be readily seen on the amplitude of the strain signals measured experimentally and as predicted by the numerical model. The amplitude of the experimentally measured strain signal can be seen to diverge from the numerical result with increasing time, while the numerical results do not show such a decrease. It should be remembered that a sandwich beam is a complex system and has various sources of damping and nonlinearity which are not included in the model. In the analysis it has been assumed that the sandwich core is perfectly elastic which is not strictly true. Another source of discrepancy between the two results is the boundary conditions used for the experiments. The beam is simply supported by fixing it between two rollers mounted on a frame attached to the vibration table. The mounting of the beam between the two rollers causes local deformation while also partially restricting the rotation of the beam around the rollers. Thus, the boundary condition is more complicated than a simple support, which causes reflections to be different from that predicted by the simply supported assumption.

For the beam with resonators, the maximum strain measured experimentally is slightly greater than that obtained numerically. For the simulations, it is assumed that all the local resonators resonate at the expected resonance frequencies. However, in reality this is not the case. Due to slight variation in the individual spring stiffness, and due to the possible friction between the copper mass and the casing interior, the resonance frequencies of the local resonators are spread over a range of frequencies close to the targeted local resonance frequency which reduces their total effectiveness.

Figures 19 and 20 compare the strains obtained experimentally and numerically, respectively. The strains obtained for the beams with resonators and with fixed masses are significantly lower than the strain for the beam without any added mass. This can be explained by the additional inertia of the heavier beams. There is a small reduction in strain for the beam with resonators as compared to the beam with fixed masses; however the reduction is not as significant as obtained by the numerical results in the previous section. As explained earlier, this is due to imperfections in the local resonators coupled with the additional energy input into the beam with t resonators as can be seen from the frequency spectrum of the input forces.

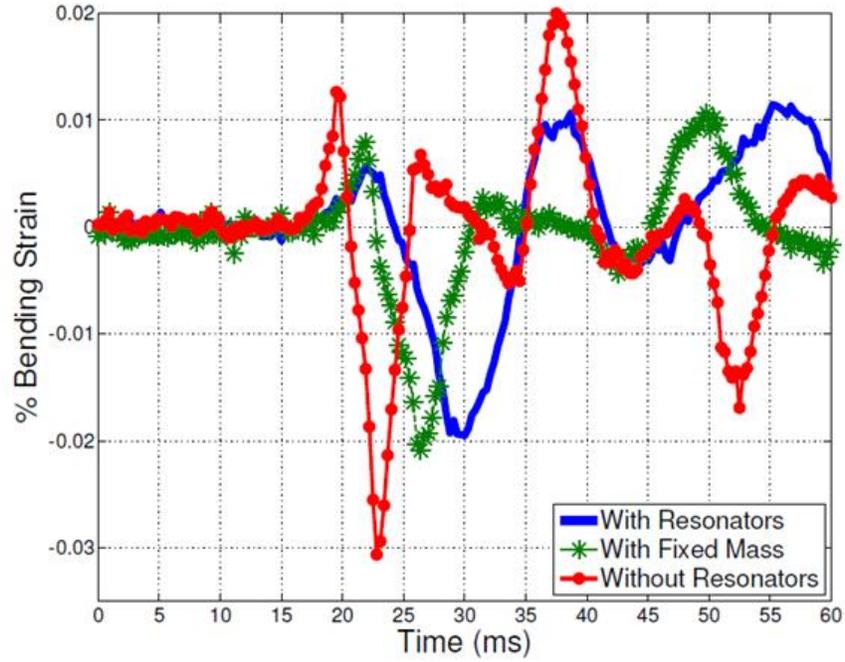

**Figure 19.** Comparison of strains measured experimentally.

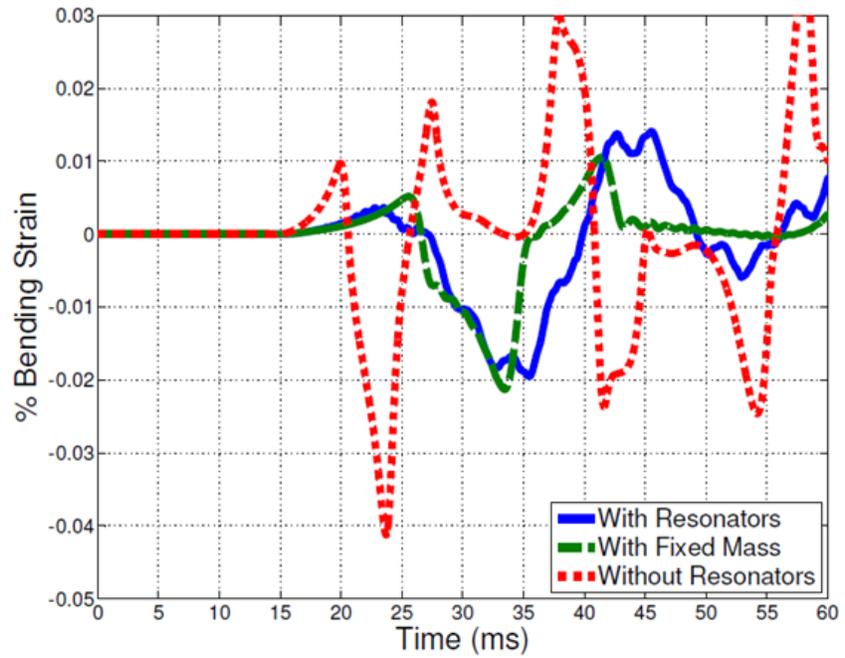

**Figure 20.** Comparison of strains obtained by finite element simulations.

In order to obtain a better comparison between the impact wave behaviors of the three beams, simulations are performed for the three beams using the same input force for all three. The impact force used is the force measured experimentally for the beam with resonators. Figure 21 shows the strains obtained at the same location as measured in the experiments. Under the same input force, given that all the resonators resonate at the same frequency and in absence of damping, the beam with resonators is far more effective in attenuating the flexural waves as compared to the beams without resonators.

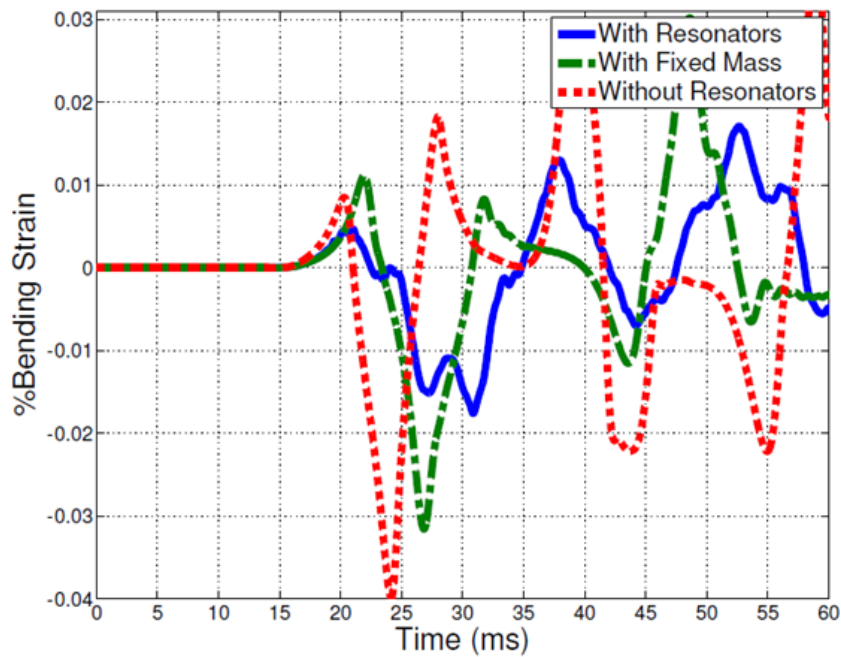

**Figure 21.** Comparison of numerically obtained strains for the three beams subjected to the same impact history.

**Conclusion**

The ability of sandwich beam with internal resonators to attenuate broad spectrum impact loads is analyzed. Finite element models are used to show that beams with resonators inserted in the core are more effective in attenuating impact loads than an equivalent beam without resonators. It is shown that the choice of the resonators depends on the impact load duration and its frequency content. For a given impact, the design of local resonators can be optimized by looking at the frequency content of the impact wave and the bandgap width generated by the local resonators. Transverse impact experiments are performed to verify the numerical results and to demonstrate the effectiveness of the resonators. A good match between the numerical results and the experiments is obtained. Thus, it is shown that sandwich beams with resonators can be used to better attenuate impact loads and for applications requiring blast load mitigation.

**Acknowledgement**

This work was supported by the Office of Naval Research [grant number N00014-11-1-0580]. Dr. Yapa D.S. Rajapakse was the program manager.